\documentclass[aps,preprintnumbers,amsmath,amssymb,nofootinbib]{revtex4}
\usepackage{graphicx}
\usepackage{color}
\usepackage{epsfig}
\def \be  {\begin{equation}}
\def \ee  {\end{equation}}
\def \ee  {\end{equation}}
\def \bea {\begin{eqnarray}}
\def \eea {\end{eqnarray}}

\def\be {\begin{equation}}
\def\ee {\end{equation}}
\def\bea {\begin{eqnarray}}
\def\eea {\end{eqnarray}}
\def\bc {\begin{center}}
\def\ec {\end{center}}
\def\bfg {\begin{figure}}
\def\efg {\end{figure}}
\def\bi {\begin{itemize}}
\def\ei {\end{itemize}}

\def\beq{\begin{equation}}
\def\eeq{\end{equation}}
\def\br{\begin{eqnarray}}
\def\er{\end{eqnarray}}

\begin{document}
\title{The Quark-Gluon Plasma Equation of State and The Generalized Uncertainty Principle}
\author{L. I. Abou-Salem}
\email{loutfy.Abou-salem@fsc.bu.edu.eg}
\affiliation{Physics Department, Faculty of Science, Benha University, Benha 13518, Egypt}

\author{N. M. El Naggar}
\email{nabile.elnagar@fsc.bu.edu.eg}
\affiliation{Physics Department, Faculty of Science, Benha University, Benha 13518, Egypt}
\author{I. A. Elmashad}
\email{ibrahim.elmashad@fsc.bu.edu.eg}
\affiliation{Physics Department, Faculty of Science, Benha University, Benha 13518, Egypt}

\begin{abstract}
\par
 The quark-gluon plasma $($QGP$)$ equation of state within a minimal length scenario or Generalized Uncertainty Principle $(GUP)$ is studied$.$ The Generalized Uncertainty Principle is implemented on deriving the thermodynamics of ideal QGP at a vanishing chemical potential$.$ We find a significant effect for the GUP term$.$ The main features of QCD lattice results were quantitatively achieved in case of $n_{f}=0$, $n_{f}=2$ and $n_{f}=2+1$ flavors for the energy density$,$ the pressure and the interaction measure$.$ The exciting point is the large value of bag pressure especially in case of $n_{f}=2+1$ flavor which reflects the strong correlation between quarks in this bag which is already expected$.$ One can notice that$,$ the asymptotic behavior which is characterized by Stephan-Boltzmann limit would be satisfied$.$
\end{abstract}
\maketitle

\hspace{+1 cm} keywords$:$ Quantum gravity$,$ Bag model$,$ Quark-gluon plasma$,$ Thermodynamics$.$
\section{Introduction}
\par
An essential modifications in the Heisenberg$'$s uncertainty principle are predicted near planck scale which is called generalized uncertainty principle GUP$.$ One of the most exciting predictions of some approaches related to quantum gravity\cite{QGPref1}$,$ perturbative string theory \cite{QGPref2}and black holes\cite{QGPref3} is the minimal length concept existence. For a recent review see \cite{QGPref4}$.$ These approaches seems to modify almost all mechanical Hamiltonians$.$ Thus, quantum mechanics can be studied in the presence of a minimal length $[5$-$ 8]$. Also$,$ in quantum optics, the GUP implications can be measured directly which confirm the theoretical predictions $[9$-$11].$ In particular, exact solutions of various relativistic $[12$-$15]$ and non relativistic problems $[16$-$22]$ has been obtained in the presence of a minimal length $\Delta x_{0}=\hbar (\beta)^{1/2}$. The generalized uncertainty principle$[23$,$24]$ was implemented on deriving the thermodynamics of ideal quark gluon plasma of massless quark flavor $[25$,$26]$.\\
\par In other minimal length formalism $[5$,$6],$ the Heisenberg algebra associated with the momentum $\hat{p}$ and the position coordinates $\hat{x_{i}}$ is given by$,$
\begin{equation}\label{1}
    [\hat{x_{i}}, \hat{p_{j}}]= i\hbar \delta_{ij} (1+\beta p^{2})
\end{equation}
Where $\beta>0$ is the minimal length parameter$.$ The generalized uncertainty principle GUP corresponding to it reads,
\begin{equation}\label{rabab121}
   \Delta x_{i}\Delta p_{j}\geq \delta_{ij}[1+\beta(\Delta p)^{2}+ \beta <p>^{2}]
\end{equation}
which yields a minimal observable length $\Delta x_{0}=\hbar (\beta)^{1/2}.$ Hence$,$ the momentum would be subject of a modification and becomes$,$
\begin{equation}\label{rabab131}
   p_{i}= p_{oi}(1+\beta p_{0}^{2})
\end{equation}
Where $x_{i}=x_{0i}$ and $p_{0i}$ satisfy the canonical commutation relations $[x_{0i},p_{0j}]= i\hbar \delta _{ij}.$ Also$,$ $p_{0i}$ can be interpreted as the momentum at low energies and $p_{i}$ at high energies$.$ Since the GUP modifies the Hamiltonian$,$ it is important to study these effects quantitatively.These effects were investigated on condensed matter$,$ atomic systems$[10$-$ 12,27],$ the liouville theorem $($LT$)$ in statistical mechanics\cite{QGPref29} and on the weak equivalence principle $(WKP).$ Recently$,$ another approach based on super gravity was implemented to QCD and to QGP especially$[29$-$ 31].$\\
\par
In this paper$,$ the effect of the GUP on QGP equation of state of massless quark flavors at a vanishing chemical potential $\mu$ is studied$.$ We calculate the corrections to various thermodynamic quantities$,$ like the energy density and the pressure$.$ Then$,$ these corrections with bag model are used to describe the quark gluon plasma equation of state and comparing it with QCD lattice results$.$
This paper is organized as follows$.$ In section $II,$ we derive thermodynamics of QGP consisting of a non-interacting massless bosons and fermions with impact of GUP approach$.$ The results, discussions and conclusions are given in sections $III$ and $IV$.
\section{Thermodynamics of Quark-Gluon Plasma with GUP effect}
In this section$,$ we derive the thermodynamics of QGP in case of bosons and fermions taking into account the GUP impact$.$ Then$,$ the thermodynamical equations such as pressure and energy density of quark gluon plasma are obtained$.$
\subsection{In case of bosons}
\par
At finite temperature $T$ and chemical potential $\mu,$ the grand-canonical partition function $z_{B}$ for non-interacting massive bosons with $g$ internal degrees of freedom is given by\cite{QGPref33}
\begin{eqnarray}\label{rabab11}
 z_{B}&=&\prod_{k} \left [\sum_{l=0}^{\infty} \exp \left(-l\dfrac{E(k)-\mu}{T}\right)\right]^{g}\\
         &=&\prod_{k} \left [1-\exp \dfrac{(E(k)-\mu)}{T}\right]^{-g}.
\end{eqnarray}
Where $l$ is the occupation number for each quantum state with energy $E(k)=\sqrt{k^{2}+m^{2}}$ with mass $m$ and $k$ is the momentum of the particle$.$ Here the infinite product is taken for all possible momentum states$.$\\
\par For simplicity we consider chiral limit $(i.e.$ vanishing mass$)$ and a vanishing chemical potential$,$ which experiments ensures at high energy$.$ Then the dominant excitation in the hadronic phase is a massless pion$,$ while that in the quark-gluon plasma is a massless quarks and gluons$.$ For a particle of mass M having a distant origin and an energy
comparable to the Planck scale$,$ the momentum would be a subject of a tiny modification and so that the dispersion relation would too$.$ According to GUP-approach$,$ the dispersion relation reads
\begin{equation}
E^{2}(k)=k^{2}c^{2}(1+\beta k^{2})^{2}+M^{2}c^{4},
\end{equation}
Where $M$ and $c$ are the mass of the particle and the speed of light as introduced by Lorentz and implemented in special relativity$,$ respectively$.$ For simplicity we use  natural units in which $\hbar=c=1$. Hence, we have
\begin{equation}
E(k)=k(1+\beta k^{2})
\end{equation}
For large volume$,$ the sum over all states of single particle can be rewritten in terms of an integral \cite{QGPref34}$.$
\begin{equation}
\sum_{k}\rightarrow{\dfrac{V}{(2\pi)^{3}}\int_{0}^{\infty}d^{3}k}\rightarrow{\dfrac{V}{2\pi^{2}}\int_{0}^{\infty}\dfrac{k^{2}dk}{(1+\beta k^{2})^{4}}}
\end{equation}
Thus$,$ the partition function$,$ eq.(\ref{rabab11})$,$ becomes
\begin{eqnarray}
\nonumber  \ln z_{B}&=& \dfrac{-V g}{2 \pi^{2}} \int_{0}^{\infty} k^{2} \dfrac{\ln \left[1-\exp\left(-\dfrac{E(k)}{T}\right)\right]}{(1+\beta k^{2})^{4}} ~dk \\
 \nonumber   \hspace{-2.75 cm}         &=& \dfrac{-V g}{2 \pi^{2}} \int_{0}^{\infty}  k^{2} \dfrac{\ln \left[1-\exp\left(-\dfrac{k}{T}(1+ \beta k^{2})\right)\right]}{(1+\beta k^{2})^{4}} ~dk.\\
  \nonumber  \hspace{-2.75 cm}         &=&\dfrac{V g}{2\pi^{2}}\left[\dfrac{k}{6\beta(1+\beta k^{2})^{3}}\ln \left[1-\exp\left(-\dfrac{k}{T}(1+\beta k^{2})\right) \right]\right] \Big{|}_{0}^{\infty} \\
 \nonumber         \hspace{-2.75 cm}            &-&\dfrac{V g}{2\pi^{2}}\int_{0}^{\infty} \dfrac{1}{6\beta(1+\beta k^{2})^{3}}\left[\ln \left[1-\exp\left(-\dfrac{k(1+\beta k^{2})}{T}\right)\right]
  +\dfrac{k(1+3\beta k^{2})}{T}\dfrac{1}{\exp\left(\dfrac{k(1+\beta k^{2})}{T}\right)-1}\right] ~dk\;\;\; \label{lnbbb}
\end{eqnarray}
Where $"$Integration by Parts$"$ was used for solving the above equation$.$ It is obvious that the first term in eq. (\ref{lnbbb}) vanishes$.$ Thus$,$
\begin{equation}
\ln z_{B}=\dfrac{-V g}{2\pi^{2}}\int_{0}^{\infty} \dfrac{1}{6\beta(1+\beta k^{2})^{3}}\left[\ln \left[1-\exp\left(-\dfrac{k(1+\beta k^{2})}{T}\right)\right]+\dfrac{k(1+3\beta k^{2})}{T}\dfrac{1}{\exp\left(\dfrac{k(1+\beta k^{2})}{T}\right)-1}\right] ~dk       \label{lnaaa}
\end{equation}
For solving eq$.$(\ref{lnaaa}) \\
Let $x=\dfrac{k}{T}(1+\beta k^{2})$ so that $dx=\dfrac{(1+3\beta k^{2})dk}{T}$ and
\begin{eqnarray}
 \nonumber   \hspace{-2.75 cm} \ln z_{B} &=& \dfrac{-V g}{2\pi^{2}}\left[\int_{0}^{\infty}\dfrac{\ln\left[1-e^{-x}\right]T dx}{6\beta(1+\beta k^{2})^{3}(1+3\beta k^{2})}\right] \\
    \hspace{-2.75 cm}                    &-& \dfrac{V g}{2\pi^{2}}\left[\int_{0}^{\infty}\dfrac{k dx}{6\beta(1+\beta k^{2})^{3}(e^{x}-1)}\right]\label{rabab555}
\end{eqnarray}
 The momentum $k$ as a function of x variable can be approximated to the first order of $\beta$ as following
\begin{eqnarray}
  k &=& xT-\beta k^{3} \\
  \nonumber   \hspace{-2.75 cm}  &=& xT-\beta (xT-\beta k^{3})^{3}\\
   \nonumber   \hspace{-2.75 cm} &=& xT-\beta x^{3}T^{3}\left(1-\dfrac{\beta k^{3}}{xT}\right)^{3}\\
    &\simeq& xT-\beta x^{3}T^{3}. \;\;\; \label{rabab26}
\end{eqnarray}
Substituting the value of k into integrant terms of eq$.$(\ref{rabab555}) and approximating them to the first order of $\beta$ as following$:$\\
For the first term$,$
\begin{eqnarray}
 \nonumber   \hspace{-2.75 cm} \dfrac{T}{6\beta (1+\beta k^{2})^{3}(1+3\beta k^{2})} &=& \dfrac{T(1-6\beta k^{2}+9\beta^{2}k^{4})}{6\beta} \\
 \hspace{-2.75 cm}  &\simeq& \dfrac{T}{6\beta}-x^{2}T^{3}+\dfrac{7\beta}{2}x^{4}T^{5} \label{rabab295}
\end{eqnarray}
For the second term$,$
\begin{eqnarray}
\nonumber \hspace{-2.75 cm} \dfrac{k}{6\beta (1+\beta k^{2})^{3}}  &=& \dfrac{k-3\beta k^{3}}{6\beta}\\
 \hspace{-2.75 cm}  &\simeq& \dfrac{xT}{6\beta}-\dfrac{2x^{3}T^{3}}{3}+\dfrac{3}{2}\beta x^{5}T^{5}\label{rabab296}
\end{eqnarray}
Substituting the values of terms in eqs$.$(\ref{rabab295}) and (\ref{rabab296}) into eq$.$(\ref{rabab555}) and solving them analytically$,$ we have
\begin{equation}\label{bosons}
\ln z_{B}=\dfrac{\pi^{2}}{90}Vg T^{3}-\dfrac{16\pi^{4}}{315}g\beta T^{5}
\end{equation}
In case of $\beta\rightarrow 0$, the above equation is reduced to the partition function for bosons without GUP effect \cite{QGPref33}.
\subsection{In case of fermions}
\par
At finite temperature $T$ and chemical potential $\mu,$ the grand-canonical partition function $z_{f}$ for non-interacting massive bosons with $g$ internal degrees of freedom is given by\cite{QGPref33}
\begin{eqnarray}\label{rabab11111}
    z_{f}&=&\prod_{k} \left [\sum_{l=0,1}^{\infty} \exp \left(-l\dfrac{E(k)-\mu}{T}\right)\right]^{g}\\
         &=&\prod_{k} \left [1+\exp \dfrac{-(E(k)-\mu)}{T}\right]^{g}.
\end{eqnarray}
Where $l$ is the occupation number for each quantum state with energy $E(k)=\sqrt{k^{2}+m^{2}}$ with mass $m$ and $k$ is the momentum of the particle$.$ Here the infinite product is taken for all possible momentum states$.$\\
\par For simplicity we consider chiral limit $(i.e.$ vanishing mass$)$ and a vanishing chemical potential$,$ which experiments ensures at high energy$.$ Then the dominant excitation in the hadronic phase is a massless pion, while that in the quark$-$gluon plasma is a massless quarks and gluons$.$ For a particle of mass M having a distant origin and an energy
comparable to the Planck scale$,$ the momentum would be a subject of a tiny modification. According to GUP-approach$,$ the dispersion relation reads
\begin{equation}
E^{2}(k)=k^{2}c^{2}(1+\beta k^{2})^{2}+M^{2}c^{4},
\end{equation}
Where $M$ and $c$ are the mass of the particle and the speed of light as introduced by Lorentz and implemented in special relativity$,$ respectively$.$ For simplicity we use  natural units in which $\hbar=c=1$. Hence, we have
\begin{equation}
E(k)=k(1+\beta k^{2})
\end{equation}
For large volume$,$ the sum over all states of single particle can be rewritten in terms of an integral \cite{QGPref34}$.$
\begin{equation}
\sum_{k}\rightarrow{\dfrac{V}{(2\pi)^{3}}\int_{0}^{\infty}d^{3}k}\rightarrow{\dfrac{V}{2\pi^{2}}\int_{0}^{\infty}\dfrac{k^{2}dk}{(1+\beta k^{2})^{4}}}
\end{equation}
Thus$,$ the partition function$,$ eq$.$(\ref{rabab11111})$,$ becomes
\begin{eqnarray}
 \nonumber \ln z_{f}&=& \dfrac{V g}{2 \pi^{2}} \int_{0}^{\infty}\dfrac{k^{2}\ln \left(1+\exp\left(-\dfrac{E(k)}{T}\right)\right)}{(1+\beta k^{2})^{4} dk}\\
 \nonumber   \hspace{-2.75 cm}  &=&\dfrac{V g}{2 \pi^{2}} \int_{0}^{\infty}\dfrac{k^{2}\ln \left(1+\exp\left(-\dfrac{k(1+\beta k^{2})}{T}\right)\right)}{(1+\beta k^{2})^{4} dk} \\
  \nonumber   \hspace{-2.75 cm} &=& \dfrac{-Vg}{2\pi^{2}}\dfrac{k\ln \left(1+\exp\left(-\dfrac{k(1+\beta k^{2})}{T}\right)\right)}{6\beta(1+\beta k^{2})^{3}}\Big{|}_{0}^{\infty}\\
  &+& \dfrac{Vg}{2\pi^{2}}\int_{0}^{\infty}\dfrac{\ln \left(1+\exp\left(-\dfrac{k(1+\beta k^{2})}{T}\right)\right) dk}{6\beta(1+\beta k^{2})^{3}} -\dfrac{Vg}{2\pi^{2}}\int_{0}^{\infty}\dfrac{k(1+3\beta k^{2}) dk}{6\beta T(1+\beta k^{2})^{3}\left(1+\exp\left(\dfrac{k(1+\beta k^{2})}{T}\right)\right)}.\label{roby852}
\end{eqnarray}
Where $"$Integration by Parts$"$ was used for solving the above equation$.$ It is obvious that the first term in eq. (\ref{roby852}) vanishes$.$ Thus$,$
\begin{equation} \label{roby2201}
\ln z_{f}=\dfrac{Vg}{2\pi^{2}}\left[\int_{0}^{\infty}\dfrac{\ln \left(1+\exp\left(-\dfrac{k(1+\beta k^{2})}{T}\right)\right) dk}{6\beta(1+\beta k^{2})^{3}}-\int_{0}^{\infty}\dfrac{k(1+3\beta k^{2}) dk}{6\beta T(1+\beta k^{2})^{3}\left(1+\exp\left(\dfrac{k(1+\beta k^{2})}{T}\right)\right)}\right]
\end{equation}
For solving eq.(\ref{roby2201}) \\
Let $x=\dfrac{k}{T}(1+\beta k^{2})$~ so that $dx=\dfrac{(1+3\beta k^{2})dk}{T}$ and
\begin{equation}\label{rabab5558}
\ln z_{f} = \dfrac{V g}{2\pi^{2}}\left[\int_{0}^{\infty}\dfrac{\ln\left[1+e^{-x}\right]T dx}{6\beta(1+\beta k^{2})^{3}(1+3\beta k^{2})}-\int_{0}^{\infty}\dfrac{k(1+3\beta k^{2})dx}{6\beta(1+3\beta k^{2})(e^{x}+1)}\right]
\end{equation}
Substituting the value of $k$ into integrant terms of eq$.$(\ref{rabab5558}) and approximating them to the first order of $\beta$ as following$:$\\
For the first term$,$
\begin{eqnarray}
 \nonumber   \hspace{-2.75 cm} \dfrac{T}{6\beta (1+\beta k^{2})^{3}(1+3\beta k^{2})} &=& \dfrac{(1-3\beta k^{2})(1-3\beta k^{2})}{6\beta} \\
 \hspace{-2.75 cm}  &\simeq& \dfrac{T}{6\beta}-x^{2}T^{3}+\dfrac{7\beta}{2}x^{4}T^{5} \label{rabab2950}
\end{eqnarray}
For the second term$,$
\begin{eqnarray}
\nonumber   \hspace{-2.75 cm} \dfrac{k}{6\beta (1+3\beta k^{2})} &=& k(1-3\beta k^{2}) \\
 \hspace{-2.75 cm}  &\simeq& \dfrac{xT}{6\beta}-\dfrac{2}{3} x^{3}T^{3}+ \dfrac{3}{2} \beta x^{5}T^{5}.\label{rabab2906}
\end{eqnarray}
Substituting the values of terms in eqs$.$(\ref{rabab2950}) and (\ref{rabab2906}) into eq$.$(\ref{rabab5558}) and solving them analytically$,$ we have
\begin{equation}\label{fermions}
\ln z_{f}=\dfrac{7}{8}\dfrac{\pi^{2}}{90}Vg T^{3}-\dfrac{31\pi^{4}}{630}Vg\beta T^{5}
\end{equation}
In case of $\beta\rightarrow 0$, the above equation is reduced to the partition function for fermions without GUP effect \cite{QGPref33}.\\
Now, the QGP equation of state of free massless quarks and gluons can be derived from the above equations.The total grand canonical partition function of QGP state can be given by adding the grand partition functions coming from the contribution of bosons$(gluons)$,fermions$(quarks)$and vacuum \cite{QGPref35} as following
\begin{equation}\label{partition}
   \ln z_{QGP}= \ln z_{F}+\ln z_{B}+\ln z_{V}
\end{equation}
Where $\ln z_{B}$, $\ln z_{F}$ and $\ln z_{V}$ are the grand canonical functions of gluons$,$ quarks and vacuum respectively$.$\\
In our case$,$ the vacuum pressure can be represented with the bag constant $B$ in case of Bag model\cite{QGPref36}$.$\\
Since$,$ the value of vacuum partition function equals to $\ln z_{V}=\dfrac{-VB}{T}$ and $\ln z=\dfrac{-\Omega}{T},$using these values eq$.$(\ref{partition}) reads
\begin{equation}\label{qgp66661}
\dfrac{\Omega}{V}\Big{|}_{QGP}=\dfrac{\Omega}{V}\Big{|}_{quarks}+\dfrac{\Omega}{V}\Big{|}_{gluons}+\dfrac{\Omega}{V}\Big{|}_{vacuum}
\end{equation}
Where $\Omega$ is the grand canonical potential. Substituting eqs. (\ref{bosons}) and (\ref{fermions}) into eq.(\ref{qgp66661})$,$ we have
\begin{equation}\label{qgp6666}
\dfrac{\Omega}{V}\Big{|}_{QGP}=-(g_{g}+\dfrac{7}{8} g_{q})\dfrac{\pi^{2}}{90}T^{4}+(g_{g}\dfrac{16\pi^{4}}{315}+g_{q}\dfrac{31\pi^{4}}{630})\beta T^{6}+B
\end{equation}
Hence$,$ the QGP equation of state reads
\begin{eqnarray}\label{qgp7766}
 \nonumber \hspace{-2.75 cm} P_{QGP}   &=&-\dfrac{\Omega}{V}\Big{|}_{QGP}\\
 \nonumber \hspace{-2.75 cm}&=&(g_{g}+\dfrac{7}{8} g_{q})\dfrac{\pi^{2}}{90}T^{4}-(g_{g}\dfrac{16\pi^{4}}{315}+
 g_{q}\dfrac{31\pi^{4}}{630})\beta T^{6}-B\\
\hspace{-2.75 cm}           &=&\dfrac{\sigma_{SB}}{3}T^{4}-\left(g_{g}\dfrac{16\pi^{4}}{315}+g_{q}\dfrac{31\pi^{4}}{630}\right)\beta T^{6}-B.\label{qgp7766}
\end{eqnarray}
Where $\sigma_{SB}$ is SB constant and is given by $\left(g_{g}+\dfrac{7}{8}g_{q}\right)\dfrac{\pi^{2}}{30}.$
Hence$,$ the energy density of QGP state of matter is given by
\begin{equation}\label{qgpenergy}
 \hspace{-2.75 cm} \varepsilon_{QGP} =\sigma_{SB}T^{4}-3\left(g_{g}\dfrac{16\pi^{4}}{315}+
g_{q}\dfrac{31\pi^{4}}{630}\right)\beta T^{6}+B.
\end{equation}
\section{Results and discussions}
\par
The main features of QCD lattice results shows that a clear $N_{f}$- dependance for both energy density and pressure $($i.e. they become larger as the number of degrees of freedom increases$)$. As it is clear from the Monte Carlo lattice results\cite{QGPref37, QGPref371}, the pressure $P(T)$ rapidly increases at $T\simeq T_{c}$ which may be agreed with our predictions after adding the GUP effect on QGP equation of state$.$ The bag pressure $B$ can be determined by comparing the obtained equation with that of QCD lattice results$.$\\
\par
The main problem appears when one start to adjust the behavior of the energy density$;$ through varying the value of the bag pressure$;$ with the QCD lattice results$,$ One can not obtain a qualitative agreement in case of the pressure using the same bag parameter value\cite{QGPref38}$.$\\
\par
For overcoming this problem$,$ the bag model was modified \cite{QGPref39}$.$ In this technique$,$ the thermodynamical relation between the energy density and the pressure\cite{QGPref34} was used.
\begin{equation}\label{relation}
T\dfrac{dP}{dT}- P(T)= \varepsilon (T)
\end{equation}
Solving the above first order differential equation$,$ we obtain the QGP equation of state as follows
\begin{equation}\label{pressure52}
P =\dfrac{\sigma_{SB}}{3}T^{4}-\left(g_{g}\dfrac{16}{525}+g_{q}\dfrac{31}{1050}\right)\pi^{4}\beta T^{6}-B+ A T
\end{equation}
Where $A$ is a constant coming from the partial differential equation solution and can be determined from comparing the calculated QGP equation of state with the QCD lattice results$.$\\
\par To adjust the high temperature behavior for $P(T)$ and $\varepsilon (T),$ we will consider the suppression factor of the
Stefan Boltzmann constant used in quasi-particle approach\cite{QGPref40}$.$In this approach, the system of interacting gluons may be treated as a non-interacting quasi particles gas with gluon quantum numbers$,$ but with thermal mass $(i.e. m(T)).$ The modified SB constant $\sigma$ reads\cite{QGPref38}$.$
\begin{equation}\label{sigma15}
\sigma= \kappa(a) \sigma_{SB}
\end{equation}
Where $\kappa(a)$ is a suppression factor$.$ For $a\rightarrow 0,$ it follows $\kappa\rightarrow 1.$ Also$,$ The function $\kappa(a)$ decreases monotonously and goes to zero at $a\rightarrow \infty.$ Thus$,$ the final form of the bag model equation of state (\ref{pressure52}) with incorporating the GUP modification is given by
\begin{equation}\label{pressure5069}
P =\dfrac{\sigma}{3}T^{4}-\left(g_{g}\dfrac{16}{525}+g_{q}\dfrac{31}{1050}\right)\pi^{4}\beta T^{6}-B+ A T
\end{equation}
With $\sigma$ and $B$ being free model parameters.
\begin{equation}\label{qgpenergy253}
 \hspace{-2.75 cm} \varepsilon =\sigma T^{4}-3\left(g_{g}\dfrac{16}{315}+
g_{q}\dfrac{31}{630}\right)\pi^{4}\beta T^{6}+B.
\end{equation}
\par
In case of $\beta\rightarrow 0$, the above equations(\ref{pressure5069}) and (\ref{qgpenergy253}) are reduced to the pressure and the energy density equations which obtained in the modified bag model \cite{QGPref39}(i.e. without incorporating the GUP modification).
\par
\begin{figure}[htb!]
\includegraphics[width=7.5 cm]{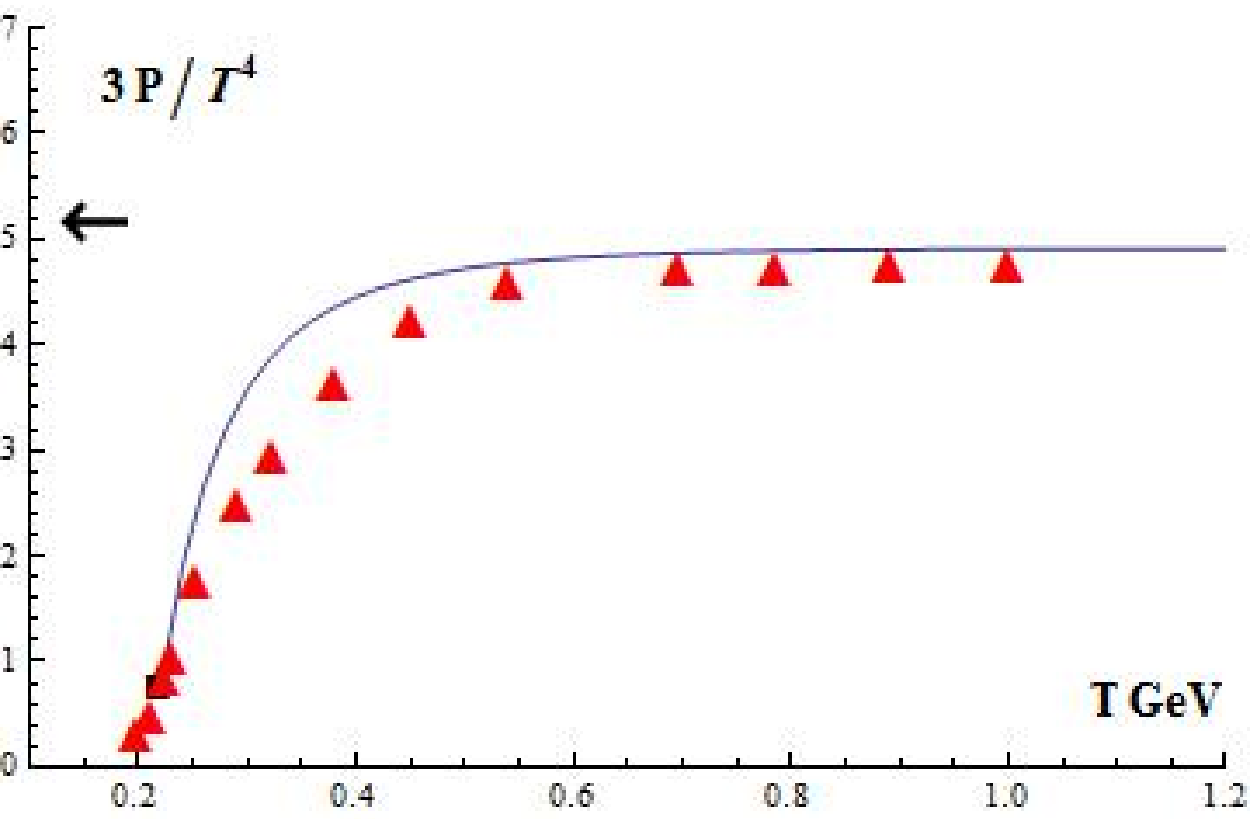}, \includegraphics[width=7.5 cm]{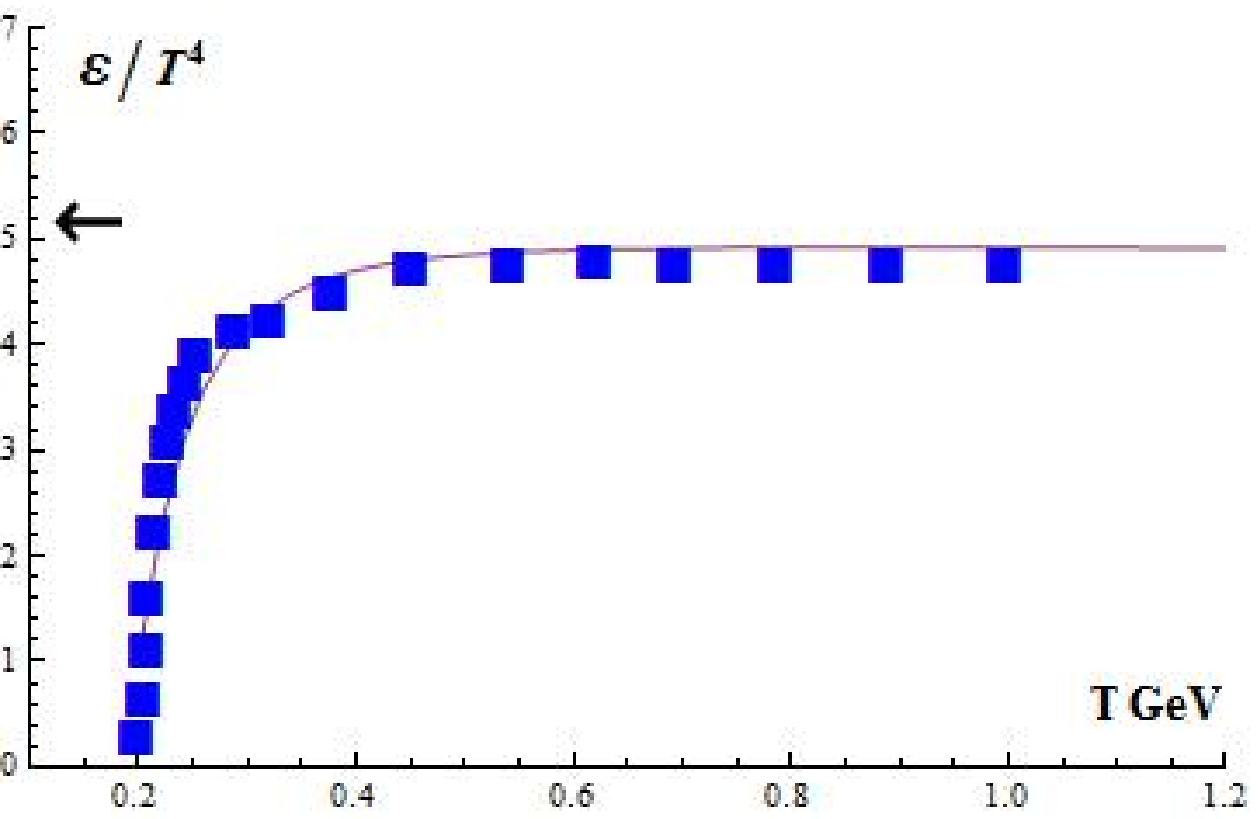}\\
\caption{ The symbols show the MC LR for the pressure and the energy density in the SU(3) gluodynamics [\cite{QGPref37}, \cite{QGPref371}], the line corresponds to the equation of state (\ref{pressure5069}) for the pressure in the left panel and (\ref{qgpenergy253}) for the energy density in the right panel  with $\sigma= 4.20719$, $A=-2.0019 T_{c}^{3}$ and $B= -1.465358 T_{c}^{4}$. The arrows in the above figures correspond to $P/\varepsilon=1/3$ (i.e. the SB limit).}\label{dalia503}
\end{figure}

The theoretical calculation of the pressure and the energy density of the QGP using eqs. (\ref{pressure5069}) and (\ref{qgpenergy253})
compared to the lattice results in case of $n_{f}=0$ are given in fig.(\ref{dalia503}). In this case we used the following parameter values; $g_{QGP}= 16, g_{q}= 0, g_{g}= 16 $, $ T_{c}= 0.200 GeV$ and $\beta = 0.0001 GeV^{-1}$ as taken in Ref. [11, 32, 39].
\par
\begin{figure}[htb!]
\includegraphics[width=7 cm]{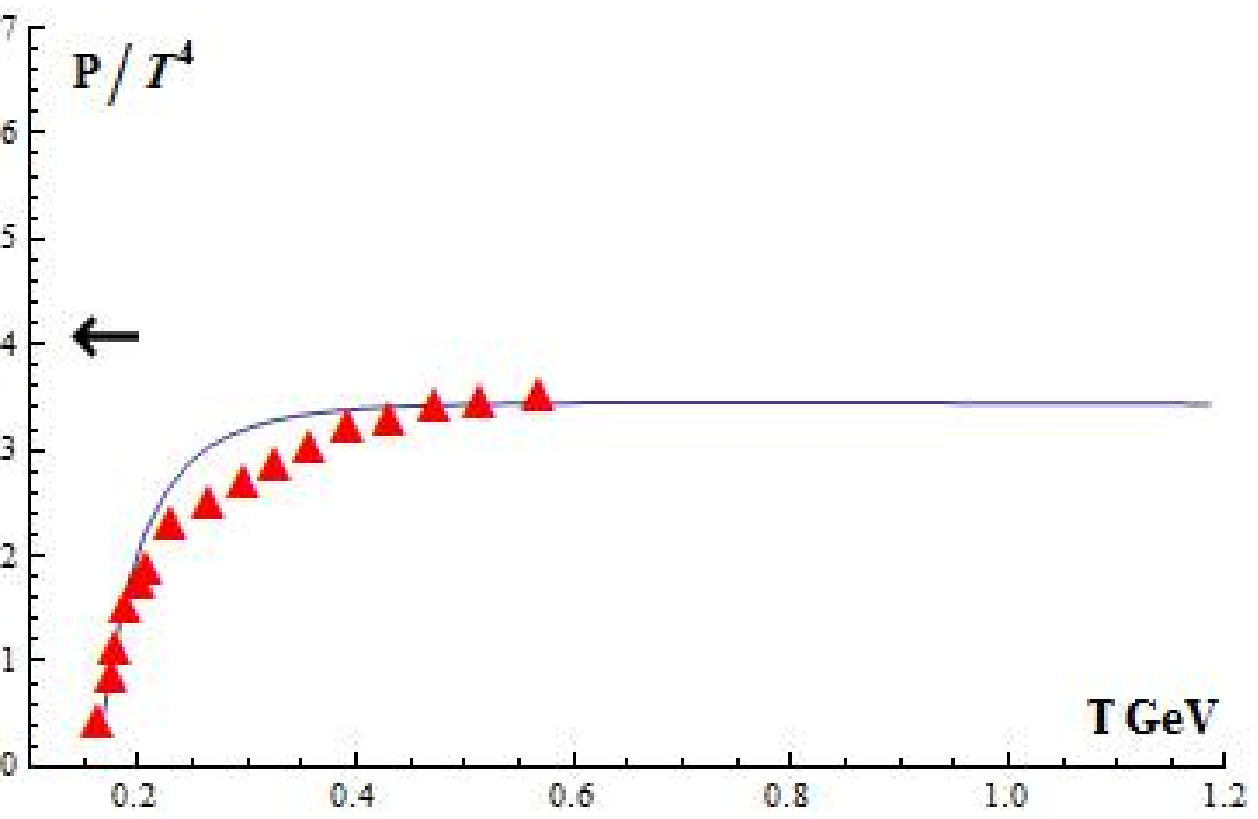}, \includegraphics[width=7 cm]{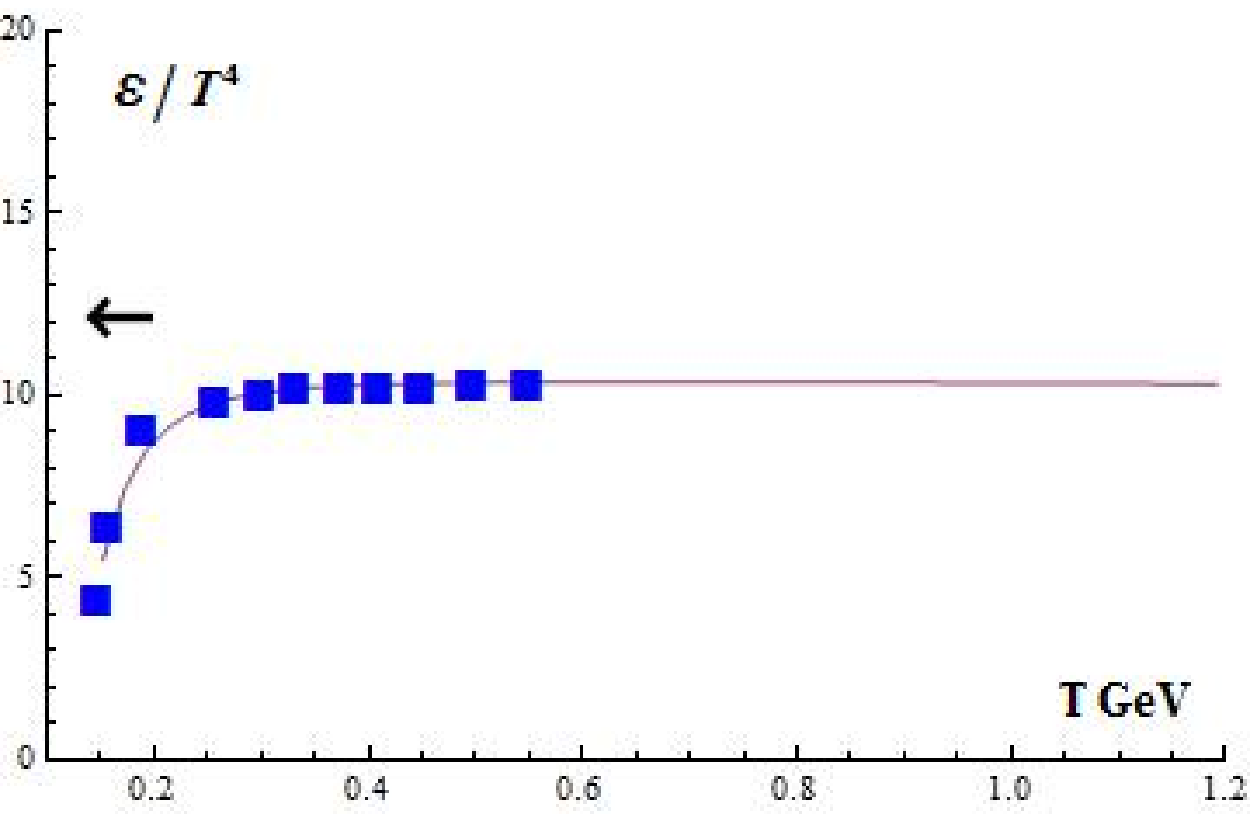}\\
\caption{ The symbols show the MC LR for the pressure and the energy density in case of $n_{f}=2$ QCD equation of state \cite{QGPref50}, the line corresponds to the equation of state (\ref{pressure5069}) for the pressure in the left panel and (\ref{qgpenergy253}) for the energy density in the right panel  with $\sigma= 10.403$, $A= - 8.72895 T_{c}^{3}$ and $B= -4.9 T_{c}^{4}$. The arrows in the above figures correspond to $P/\varepsilon=1/3$ (i.e. the SB limit).}\label{dalia1185}
\end{figure}

The theoretical calculation of the pressure and the energy density of the QGP using eqs. (\ref{pressure5069}) and (\ref{qgpenergy253})
compared to the lattice results in case of $n_{f}= 2$ are given in fig.(\ref{dalia1185}). In this case we used the following parameter values; $g_{QGP}= 37, g_{q}= 24, g_{g}= 16 $, $ T_{c}= 0.152 GeV$ and $\beta = 0.0001 GeV^{-1}$ as taken in Ref. [11, 32, 39].
\par
\begin{figure}[htb!]
\includegraphics[width=7 cm]{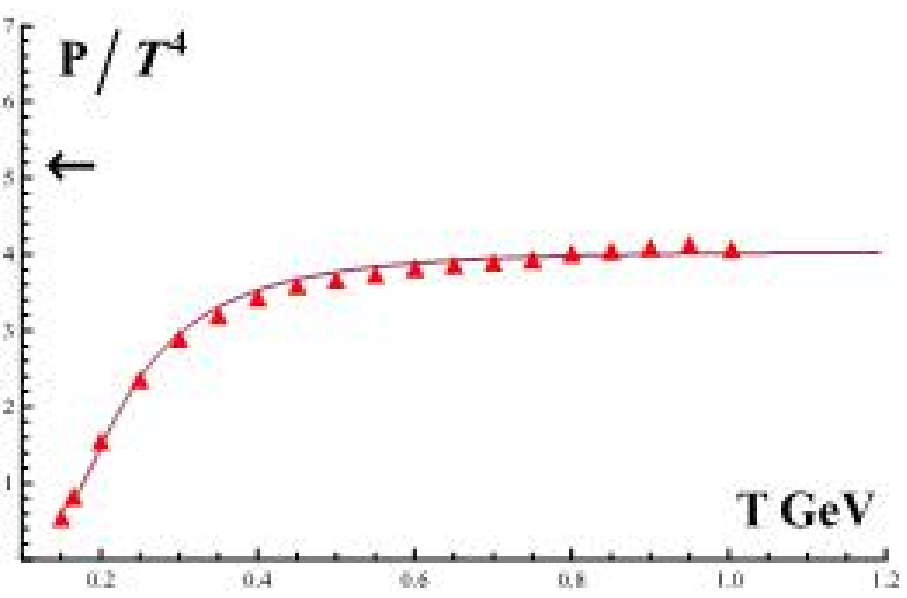}, \includegraphics[width=7 cm]{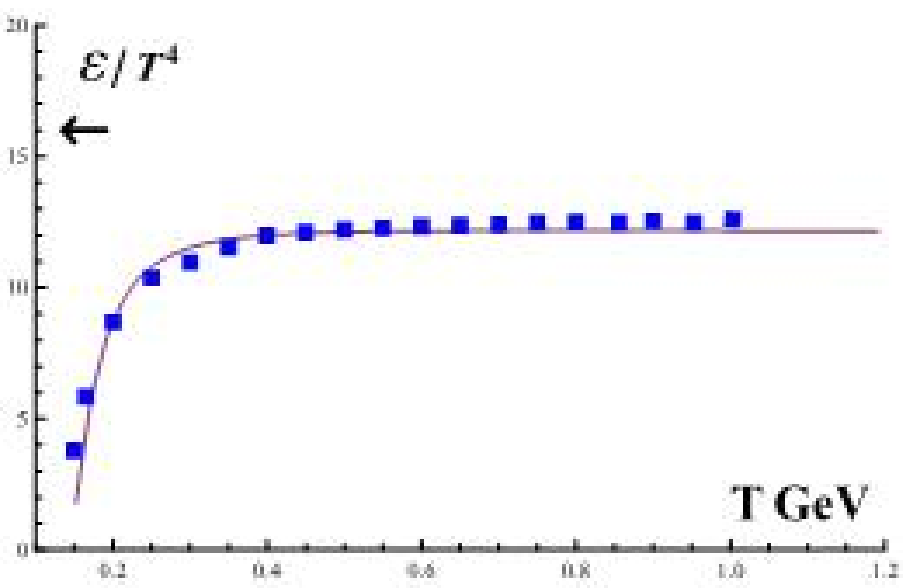}\\
\caption{ The symbols show the MC LR for the pressure and the energy density in case of $n_{f}=2+1$ QCD equation of state \cite{QGPref45}, the line corresponds to the equation of state (\ref{pressure5069}) for the pressure in the left panel and (\ref{qgpenergy253}) for the energy density in the right panel  with $\sigma= 12.22017$, $A=-13.8577 T_{c}^{3}$ and $B= -10.4353 T_{c}^{4}$. The arrows in the above figures correspond to $P/\varepsilon=1/3$ (i.e. the SB limit).}\label{dalia1152}
\end{figure}

The theoretical calculation of the pressure and the energy density of the QGP using eqs. (\ref{pressure5069}) and (\ref{qgpenergy253})compared to the lattice results in case of $n_{f}= 2+1$ are given in fig.(\ref{dalia1152}). In this case we used the following parameter values; $g_{QGP}= 47.5, g_{q}= 36, g_{g}= 16 $, $ T_{c}= 0.152 GeV$ and $\beta = 0.0001 GeV^{-1}$ as taken in Ref. $[11, 32, 39]$.\\


\begin{figure}[htb!]
\includegraphics[width=7.5 cm]{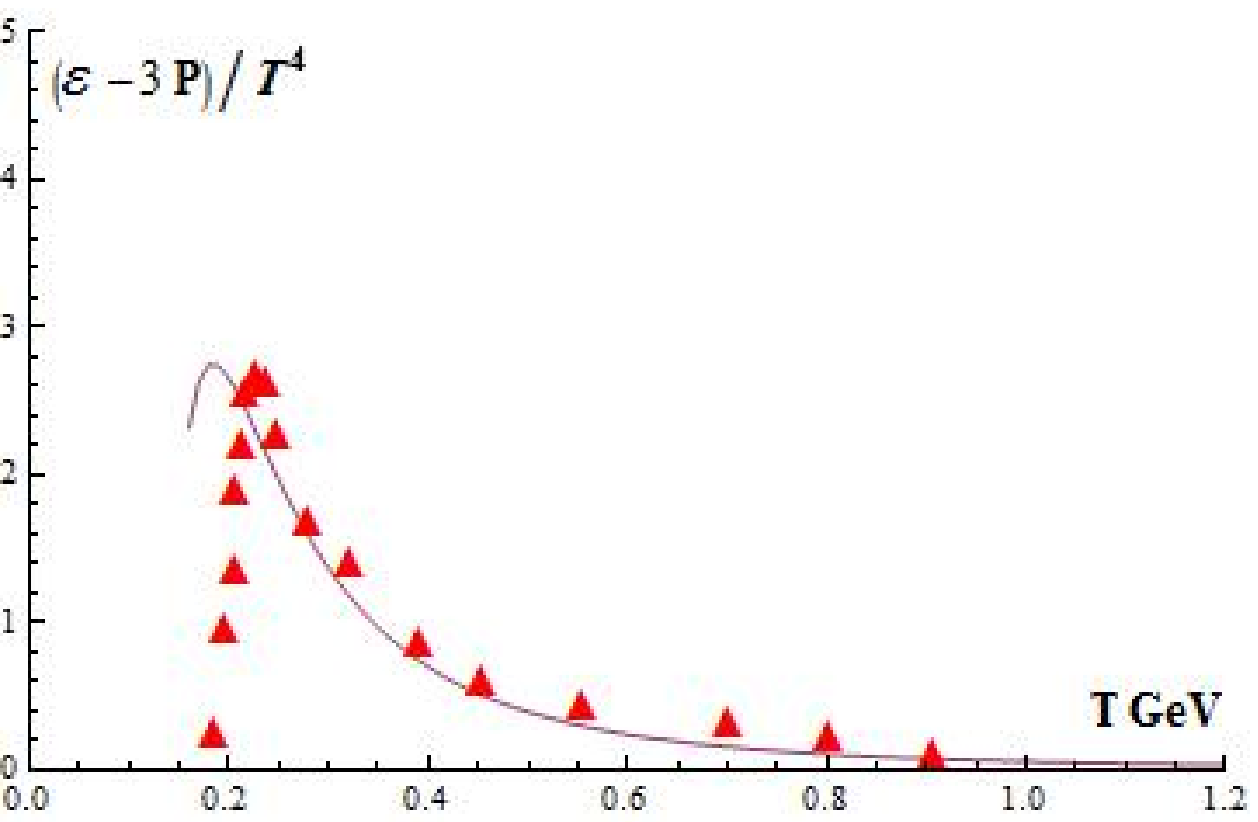}, \includegraphics[width=7.5 cm]{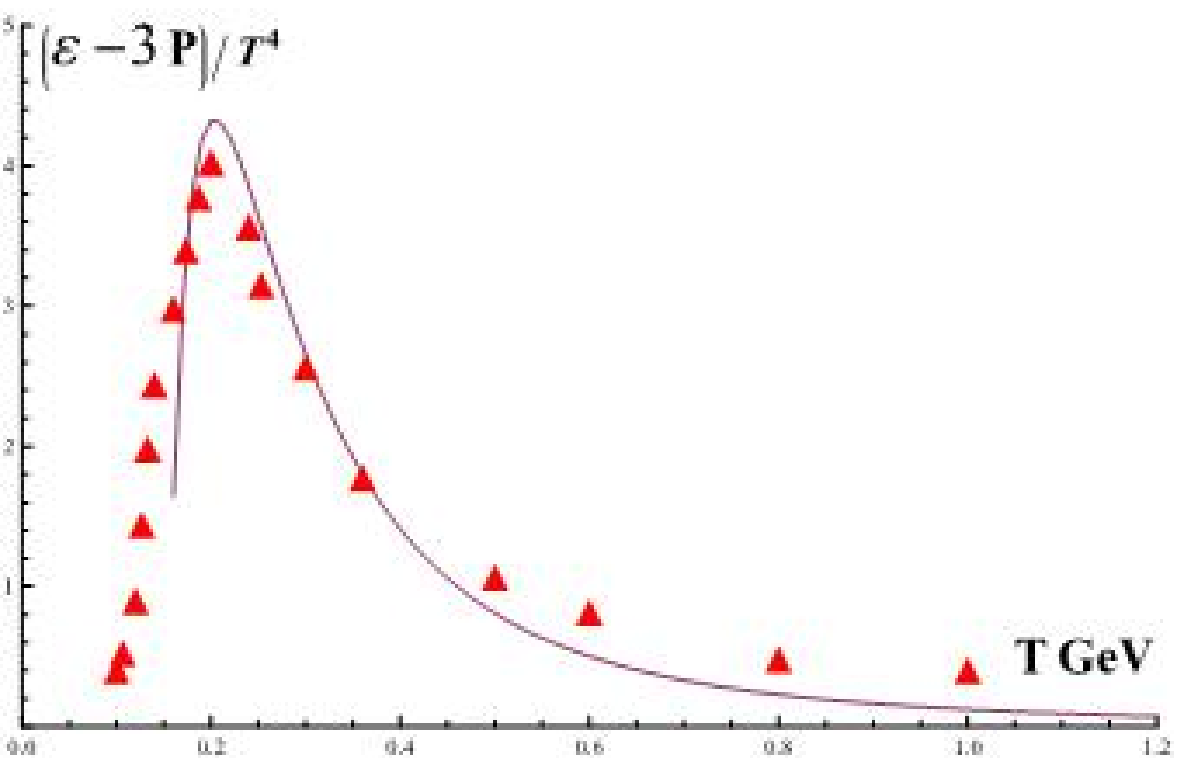}\\
\caption{ The symbols show the lattice results for the interaction measure in the $SU(3)$ gluodynamics (left panel)\cite{QGPref37} and in case of $n_{f}= 2+1$ QCD equation of state(right panel)\cite{QGPref45}}\label{dalia1155}
\end{figure}

Also$,$ we can calculate the interaction measure $(\varepsilon-3P)/T^{4}$ for both $n_{f}= 0$, and $n_{f}= 2+1$ and comparing them with lattice results for the interaction measure in the SU$(3)$ gluodynamics \cite{QGPref37} and in case of $n_{f}= 2+1$ QCD equation of state \cite{QGPref45} respectively. These results can be shown in fig$.$(\ref{dalia1155}). It can be shown that the inflection point
of $(\varepsilon-3P)/T^{4}$ is at $T=0.1846 GeV$ for $n_{f}=0$ and $T=0.202 GeV$ for $n_{f}=2+1$ which are different from that obtained in \cite{QGPref39}.  .

\newpage
\section{Conclusion}
\par
In the present work, the effect of the GUP on QGP  of massless quark flavors at a vanishing chemical potential,$\mu$, is studied. The equation of state and the energy density are derived for the QGP state consisting of a non-interacting massless bosons and fermions with impact of GUP approach.  Also, the total grand canonical partition function of QGP state is given.   One can conclude that, a significant effect for the GUP term exists in case of study the thermal properties of the QGP state. The main features of QCD lattice results were quantitatively achieved in case of $n_{f}= 0$, $n_{f}=2$ and $n_{f}=2+1$ for the equation of state, the energy density and the interaction measure. The interesting point in our results is the large value of bag pressure especially in case of $n_{f}=2+1$ flavor. It nearly equals to $4.46$ times of the value obtained in Ref.\cite{QGPref39} without taking into account the negative sign of it, which reflects the strong correlation between quarks inside the bag which is already expected. The negative sign may be regarded as the tendency of the bag to reduce its volume. One can conclude that, the modification of the QGP bag model equation of state using the GUP effect can reproduce the QCD lattice results which stands for the real data of QGP. Finally, with this study, one may be encouraged to implement GUP effect in  other thermodynamic properties of QGP.


\end{document}